\documentstyle[twoside]{article}
\catcode`\@=11
\long\def\@makefntext#1{
\protect\noindent \hbox to 3.2pt {\hskip-.9pt  
$^{{\eightrm\@thefnmark}}$\hfil}#1\hfill}

\def\@makefnmark{\hbox to 0pt{$^{\@thefnmark}$\hss}}
\def\ps@myheadings{\let\@mkboth\@gobbletwo
\def\@oddhead{\hbox{}
\rightmark\hfil\eightrm\thepage}   
\def\@oddfoot{}\def\@evenhead{\eightrm\thepage\hfil
\leftmark\hbox{}}\def\@evenfoot{}
\def\sectionmark##1{}\def\subsectionmark##1{}}
\oddsidemargin=\evensidemargin
\newcounter{sectionc}\newcounter{subsectionc}\newcounter{subsubsectionc}
\renewcommand{\section}[1] {\vspace{12pt}\addtocounter{sectionc}{1} 
\setcounter{subsectionc}{0}\setcounter{subsubsectionc}{0}\noindent 
 {\tenbf\thesectionc. #1}\par\vspace{5pt}}
\renewcommand{\subsection}[1]
{\vspace{12pt}\addtocounter{subsectionc}{1} 
 \setcounter{subsubsectionc}{0}\noindent 
{\bf\thesectionc.\thesubsectionc. {\kern1pt \bfit #1}}\par\vspace{5pt}}
\renewcommand{\subsubsection}[1]
{\vspace{12pt}\addtocounter{subsubsectionc}{1}
 \noindent{\tenrm\thesectionc.\thesubsectionc.\thesubsubsectionc.
 {\kern1pt \tenit #1}}\par\vspace{5pt}}

\newcounter{appendixc}
\newcounter{subappendixc}[appendixc]
\newcounter{subsubappendixc}[subappendixc]
\renewcommand{\thesubappendixc}{\Alph{appendixc}.\arabic{subappendixc}}
\renewcommand{\thesubsubappendixc}
 {\Alph{appendixc}.\arabic{subappendixc}.\arabic{subsubappendixc}}
\renewcommand{\appendix}[1] {\vspace{12pt}
        \refstepcounter{appendixc}
        \setcounter{figure}{0}
        \setcounter{table}{0}
        \setcounter{lemma}{0}
        \setcounter{theorem}{0}
        \setcounter{corollary}{0}
        \setcounter{definition}{0}
        \setcounter{equation}{0}
        \renewcommand{\thefigure}{\Alph{appendixc}.\arabic{figure}}
        \renewcommand{\thetable}{\Alph{appendixc}.\arabic{table}}
        \renewcommand{\theappendixc}{\Alph{appendixc}}
        \renewcommand{\thelemma}{\Alph{appendixc}.\arabic{lemma}}
        \renewcommand{\thetheorem}{\Alph{appendixc}.\arabic{theorem}}
     \renewcommand{\thedefinition}{\Alph{appendixc}.\arabic{definition}}
       \renewcommand{\thecorollary}{\Alph{appendixc}.\arabic{corollary}}
        \renewcommand{\theequation}{\Alph{appendixc}.\arabic{equation}}
        \noindent{\tenbf Appendix \theappendixc #1}\par\vspace{5pt}}
\newcommand{\subappendix}[1] {\vspace{12pt}
        \refstepcounter{subappendixc}
        \noindent{\bf Appendix \thesubappendixc. {\kern1pt \bfit #1}}
 \par\vspace{5pt}}
\newcommand{\subsubappendix}[1] {\vspace{12pt}
        \refstepcounter{subsubappendixc}
       \noindent{\rm Appendix \thesubsubappendixc. {\kern1pt \tenit #1}}
 \par\vspace{5pt}}
\newcommand{\textlineskip}{\baselineskip=13pt}
\newcommand{\smalllineskip}{\baselineskip=10pt}
\def\eightcirc{
\begin{picture}(0,0)
\put(4.4,1.8){\circle{6.5}}
\end{picture}}
\def\eightcopyright{\eightcirc\kern2.7pt\hbox{\eightrm c}} 
\newcommand{\copyrightheading}[1]
 {\vspace*{-2.5cm}\smalllineskip{\flushleft
 {\footnotesize International Journal of Modern Physics B {\bf 12} (1998) 591-600}\\
  }}
\newcommand{\pub}[1]{{\begin{center}\footnotesize\smalllineskip 
 \ \\
 \end{center}
 }}
\newcommand{\publisher}[2]{{\begin{center}\footnotesize\smalllineskip 
 \ 
 \end{center}
 }}
\def\abstracts#1#2#3{{
 \centering{\begin{minipage}{4.5in}\baselineskip=10pt\footnotesize
 \parindent=0pt #1\par 
 \parindent=15pt #2\par
 \parindent=15pt #3
 \end{minipage}}\par}}

\renewenvironment{thebibliography}[1]   
 {\frenchspacing
  \ninerm\baselineskip=11pt
  \begin{list}{\arabic{enumi}.}
 {\usecounter{enumi}\setlength{\parsep}{0pt}
  \setlength{\leftmargin 12.7pt}{\rightmargin 0pt} 
  \setlength{\itemsep}{0pt} \settowidth
 {\labelwidth}{#1.}\sloppy}}{\end{list}}
\newcounter{itemlistc}
\newcounter{romanlistc}
\newcounter{alphlistc}
\newcounter{arabiclistc}

\newcommand{\fcaption}[1]{
        \refstepcounter{figure}
        \setbox\@tempboxa = \hbox{\footnotesize Fig.~\thefigure. #1}
        \ifdim \wd\@tempboxa > 5in
           {\begin{center}
        \parbox{5in}{\footnotesize\smalllineskip Fig.~\thefigure. #1}
            \end{center}}
        \else
             {\begin{center}
             {\footnotesize Fig.~\thefigure. #1}
              \end{center}}
        \fi}
\newcommand{\tcaption}[1]{
        \refstepcounter{table}
        \setbox\@tempboxa = \hbox{\footnotesize Table~\thetable. #1}
        \ifdim \wd\@tempboxa > 5in
           {\begin{center}
        \parbox{5in}{\footnotesize\smalllineskip Table~\thetable. #1}
            \end{center}}
        \else
             {\begin{center}
             {\footnotesize Table~\thetable. #1}
              \end{center}}
        \fi}
\def\@citex[#1]#2{\if@filesw\immediate\write\@auxout
 {\string\citation{#2}}\fi
\def\@citea{}\@cite{\@for\@citeb:=#2\do
 {\@citea\def\@citea{,}\@ifundefined
 {b@\@citeb}{{\bf ?}\@warning
 {Citation `\@citeb' on page \thepage \space undefined}}
 {\csname b@\@citeb\endcsname}}}{#1}}
\newif\if@cghi
\def\cite{\@cghitrue\@ifnextchar [{\@tempswatrue
 \@citex}{\@tempswafalse\@citex[]}}
\def\citelow{\@cghifalse\@ifnextchar [{\@tempswatrue
 \@citex}{\@tempswafalse\@citex[]}}
\def\@cite#1#2{{$\null^{#1}$\if@tempswa\typeout
 {IJCGA warning: optional citation argument 
 ignored: `#2'} \fi}}

\def\pmb#1{\setbox0=\hbox{#1}
 \kern-.025em\copy0\kern-\wd0
 \kern.05em\copy0\kern-\wd0
 \kern-.025em\raise.0433em\box0}

\def\fnt#1#2{\footnotetext{\kern-.3em
 {$^{\mbox{\scriptsize #1}}$}{#2}}}
\def\fpage#1{\begingroup
\voffset=.3in
\thispagestyle{empty}\begin{table}[b]\centerline{\footnotesize #1}
 \end{table}\endgroup}
\def\runninghead#1#2{\pagestyle{myheadings}
\markboth{{\protect\footnotesize\it{\quad #1}}\hfill}
{\hfill{\protect\footnotesize\it{#2\quad}}}}
\font\tenrm=cmr10
\font\tenit=cmti10 
\font\tenbf=cmbx10
\font\bfit=cmbxti10 at 10pt
\font\ninerm=cmr9
\font\nineit=cmti9
\font\ninebf=cmbx9
\font\eightrm=cmr8

\def\qed{\hbox{${\vcenter{\vbox{   
   \hrule height 0.4pt\hbox{\vrule width 0.4pt height 6pt
   \kern5pt\vrule width 0.4pt}\hrule height 0.4pt}}}$}}

\def\bsc{{\sc a\kern-6.4pt\sc a\kern-6.4pt\sc a}} 
\def\bflatex{\bf L\kern-.30em\raise.3ex\hbox{\bsc}\kern-.14em 
T\kern-.1667em\lower.7ex\hbox{E}\kern-.125em X}

\begin{document}

\runninghead{D.~Mozyrsky, V.~Privman \& S.~P.~Hotaling}{Design of
gates for quantum computation: the XOR gate}

\normalsize\textlineskip
\thispagestyle{empty}
\setcounter{page}{1}

\copyrightheading{}  

\vspace*{0.88truein}

\fpage{1}
\centerline{\bf \hphantom{A}}
\vspace*{0.035truein}
\centerline{\bf \hphantom{A}}
\vspace*{0.035truein}
\centerline{\bf \hphantom{A}}
\vspace*{0.035truein}
\centerline{\bf DESIGN OF GATES FOR QUANTUM COMPUTATION:}
\vspace*{0.035truein}
\centerline{\bf THE THREE-SPIN XOR GATE IN TERMS OF}
\vspace*{0.035truein}
\centerline{\bf TWO-SPIN INTERACTIONS}
\vspace*{0.37truein}
\centerline{\footnotesize DIMA MOZYRSKY, \ VLADIMIR PRIVMAN}
\vspace*{0.018truein}
\centerline{\footnotesize\it Department of Physics, Clarkson University,
Potsdam, New York 13699-5820, USA} 
\vspace*{10pt}
\centerline{\normalsize and}
\vspace*{10pt}
\centerline{\footnotesize STEVEN P.~HOTALING}
\vspace*{0.016truein}
\centerline{\footnotesize\it Air Force Materiel Command, Rome
Laboratory/Photonics Division}
\baselineskip=10pt
\centerline{\footnotesize\it 25 Electronic Parkway, Rome,
New York 13441-4515, USA}
\vspace*{0.225truein}
\publisher{5 May 1997}{ }

\vspace*{0.21truein}
\abstracts{We propose to design multispin quantum gates in
which the input
and
output two-state systems (spins) are not necessarily identical.
We
describe the motivations for such studies and then derive an
explicit
general {\it two-spin interaction\/} Hamiltonian which
accomplishes
the quantum XOR gate function for a system of three spins: two
input
and one output.}{}{}

\vspace*{1pt}\textlineskip \vfill\eject
\section{Introduction} 
\noindent
Dimensions of semiconductor computer components will soon reach$^1$
about $0.25\,\mu$m $=2500\,$\AA, still well above
the sizes at which quantum-mechanical effects are important.
However, it is generally expected that as the miniaturization
continues,
atomic dimensions will be reached, perhaps, with technology
different
from today's semiconductors. The physics of quantum-mechanical
computation has attracted much
attention recently.$^2$ Quantum computer is a
quantum-coherent system that functions as
a programmable calculational apparatus,
quite unlike its classical
counterparts. It can perform certain tasks$^2$ much
faster than the classical computer: the quantum interference
property
yields$^2$ the fast-factoring (Shor's), as well as certain
other fast algorithms.
Recent theoretical results have included identification of
universal reversible
two-bit gates$^3$ and advances in error correction.$^4$
There have also been experiments and proposals for
experiments,$^5$
realizing the simplest gates.
However, the actual construction of a
macroscopic computer out of a large
number of quantum bits (two-state systems, qubits) 
is ellusive$^6$ at the present
stage of technology.
The main obstacle is the sensitivity of coherent quantum
evolution and interference
to undesirable external interactions such as noise or other
failures in
operation,$^{2,6,7}$ even though a number of error correction
schemes have
been proposed.$^4$ 

We therefore propose an alternative approach along the lines of
the
``classical'' analog computer design, of
operating the computer as a {\it single unit\/} performing in
one shot a complex logical
task instead of a network of simple gates each performing a
simple ``universal-set''
logical function. The computer as a whole
will still be subject to errors. However, these will not
be magnified by
proliferation of sub-steps each of which must be exactly
controlled.

Indeed, quantum
(and more generally reversible) computation must be externally
timed: the time scale
of the operation of each gate is determined by the interactions
rather than by the relaxation
processes as in the ordinary computer. Furthermore, gate
interactions must be externally switched$^2$
on and off because the gates affect each other's operation.
Time dependence smoother than the on/off protocol is possible: see
Ref.~8.

In fact, we consider it likely that technological advances
might first allow design
and manufacturing of limited size units, based on several tens
of atomic two-level
systems, operating in a coherent fashion over sufficiently
large time interval to
function as parts of a larger classical (dissipative) computer.
We would
like these to function as single units rather than being
composed of many gates.

While in principle in a reversible computational unit input and
output
spins (qubits) need not be different, for larger units
interacting with
the external world it may be practically useful to consider
input and output separate (or at least not identical).   
Indeed, the interactions that feed in the input
need not
necessarily be identical to those interactions/measurements
that read off the output.

We consider in this work a
{\it spatially extended\/} XOR gate based on three
spins: two input
and one output.
Generally, we have to address a complicated set of
problems: can
multispin computational units be designed with
short-range,
two-particle interactions? Can they
accomplish logical functions with interactions of the
form familiar in
condensed-matter or other experimental systems? These
and similar
questions can only be answered by multispin-unit
calculations
which will have to be numerical. Analytical results are 
limited
to the simplest gates such as NOT and XOR, the former
studied in
Ref.~8, and they provide only a partial picture.

We emphasize that the XOR function is used only as a
{\it solvable example\/} of a gate with more than two spins,
 in which one can seek
to accomplish a useful logical function {\it solely
with two-spin interactions}. 
The XOR function can be also realized with 
two spins (one of the inputs serving as the output), 
 for instance as the sub-result of the controlled-NOT 
 gate.$^{2,5,9}$ It is also
important to emphasize that while moving from two spins 
 to three spins brings in the
 issue of the two-body interactions, the other
 important aspect of
multispin gates: having the interactions {\it
short-range}, can only
 be explored with larger systems.

In Section 2, we define the
problem and introduce some notation. In Section 3, we
analyze the
matrix forms of the unitary evolution operator and
Hamiltonian
operator. The latter is explicitly calculated in
Section 4 and then
further refined in Sections 4 and 5 to yield
a two-spin-interaction result.

\section{The Three-Spin XOR Gate}
\noindent
We will use the term ``spin''
to describe a two-state system, and we will represent
spin-$1\over
2$-particle spin-components (measured in units of $\hbar /2$)
by
the standard Pauli
matrices $\sigma_{x,y,z}$. We denote by $A$, $B$, $C$ the
three two-state systems, i.e., three spins, involved. We 
are particularly interested in such initial conditions,
at time $t$, that
the input spins $A$ and $B$ are in one of the basis states
$|AB\rangle=
|11\rangle$, $|10\rangle$, $|01\rangle$, or $|00\rangle$, where
1 and
0 refer to the eigenstates of the $z$-component of the spin
operator,
with 1 referring to the ``up'' state and 0 referring to the
``down''
state. We use this convention for consistency with the
classical ``bit'' notion.
The initial state of $C$ is arbitrary.

We would like to have a quantum evolution which, provided $A$
and
$B$ are initially in those basis states, mimics the XOR
function:
\begin{equation}\matrix{A&B&{\rm
output}\cr{}1&1&0\cr{}1&0&1\cr{}0&1&1\cr{}0&0&0}
\end{equation}
where the output is at time $t+\Delta t$. As already
mentioned, one way to accomplish
this is to produce the output in $A$ or $B$, i.e.,
work with a two-spin system.
The Hamiltonian for such a system is not unique. Explicit
examples
can be found in Refs.~2, 5, 9. In the case of two spins
involved, the interactions can be single- and two-spin only.

An important question is whether multispin systems can produce
useful
logical operations with only two-spin and, for larger systems,
short-range interactions. Indeed,
two-particle short-range
interactions are much better studied and accessible to
experimental probe than multiparticle interactions. As a
solvable example that addresses the former issue, here we
 require that the XOR result be put in $C$ at time
$t+\Delta t$. The final states of $A$ and $B$, as well
as the {\it phase\/} of $C$ are arbitrary. In
fact, there are many different unitary transformations, $U$,
that correspond to the desired evolution in the eight-state
space with
the basis $|ABC\rangle=|111\rangle$, $|110\rangle$,
$|101\rangle$,
$|100\rangle$, $|011\rangle$, $|010\rangle$, $|001\rangle$,
$|000\rangle$, which we will use in this order. The choice of
the
transformation determines what happens when the initial state
is a
superposition of the reference states, what are the phases in
the output, etc.

Let us consider first, for illustration, the following
Hamiltonian,
which elludes to our more general results below,
\begin{equation} H={\pi \hbar \over 4 \Delta t}
 \left( \sqrt{2} \sigma_{zA} \sigma_{yB}
+ \sqrt{2} \sigma_{zB} \sigma_{yC}-\sigma_{yB} \sigma_{xC}
\right) \, , \end{equation}
It is written here in terms of the spin components; the
subscripts $A,B,C$ denote the spins. In the
eight-state basis specified earlier, its matrix can be obtained
by
direct product of the Pauli matrices and unit $2\times 2$
matrices
$\cal I$. For
instance, the first interaction term is proportional to
\begin{equation} \sigma_{zA}\otimes\sigma_{yB}\otimes{\cal 
I}_C \, , \end{equation}
etc. This Hamiltonian involves only two-spin-component
interactions.
In fact, in this particular example $A$ and $C$ only interact
with $
B$.

One can show that the Hamiltonian (2) corresponds to the XOR
result
in $C$ at $t+\Delta t$ provided $A$ and $B$ where in one of the
superpositions of the appropriate ``binary'' states at
$t$
(we refer to superposition here because $C$ is arbitrary at
$t$).
There are two ways to verify this. Firstly, one can diagonalize
$H$ and then calculate the unitary evolution operator
(matrix) $U$ in the diagonal representation by using
the general relation (valid for Hamiltonians which are constant
during the time interval $\Delta t$; see Ref.~8 for a
formulation that
introduces a multiplicative time dependence
in $H$),
\begin{equation} U=\exp \left(-iH\Delta t /\hbar \right)\,,
\end{equation}
and then reverse the diagonalizing transformation. The
calculation is quite cumbersome.

The second, more general approach adopted here is to
``design'' a whole family of two-spin-interaction
Hamiltonians of which the form (2) is but a special case, by
analyzing generally a family of $8\times 8$ unitary
matrices
corresponding to the XOR evolution.
This ``design'' program is carried out in the
following sections.

\section{The Structure of the Unitary Matrix and Hamiltonian}
\noindent
We require any linear combination of
the states $|\underline{11}1\rangle$ and
$|\underline{11}0\rangle$ to evolve
into a linear combination of $|11\underline{0}\rangle$,
$|10\underline{0}\rangle$, $|01\underline{0}\rangle$, and
$|00\underline{0}\rangle$, compare the underlined quantum
numbers with
the first entry in (1), with similar rules for the other
three entries in (1).

In the matrix notation, and in the standard basis introduced
earlier, namely,
$|ABC\rangle=|111\rangle$, $|110\rangle$, $|101\rangle$,
$|100\rangle$, $|011\rangle$, $|010\rangle$, $|001\rangle$,
$|000\rangle$, the most general XOR evolution operator
corresponding
to the Boolean function (1), with the output in $C$, is,
therefore
\begin{equation} U=\pmatrix{0&0&U_{13}&U_{14}&U_{15}&U_{16}&0&0\cr
 U_{21}&U_{22}&0&0&0&0&U_{27}&U_{28}\cr
 0&0&U_{33}&U_{34}&U_{35}&U_{36}&0&0\cr
 U_{41}&U_{42}&0&0&0&0&U_{47}&U_{48}\cr
 0&0&U_{53}&U_{54}&U_{55}&U_{56}&0&0\cr
 U_{61}&U_{62}&0&0&0&0&U_{67}&U_{68}\cr
 0&0&U_{73}&U_{74}&U_{75}&U_{76}&0&0\cr
 U_{81}&U_{82}&0&0&0&0&U_{87}&U_{88}}\,. \end{equation}
The condition of unitarity, $UU^\dagger=1$, reduces the
number of
independent parameters. Still they are too numerous
for the problem to be
manageable analytically; recall that each nonzero element
$U_{kn}$ is
complex and therefore involves two real parameters. Thus, we
are going
to consider a subset of operators of the form (5).

{}From our earlier work$^8$ we know that one convenient way to
reduce
the number of parameters and at the same time ensure
unitarity is to have a single phase factor in each column and
row of
the matrix. Furthermore, we choose a form which is diagonal in
the
states of the $A$-spin,
\begin{equation} U=\pmatrix{V_{4\times 4}&0_{4\times 4}\cr
0_{4\times 4}&W_{4\times 4}}\,. \end{equation}
Thus, $A$ and $B$ are not treated symmetrically.
Here $0_{4\times 4}$ denotes the $4\times 4$ matrix of zeros.
The $4\times 4$ matrices $V$ and $W$ are parametrized as
follows:
\begin{equation} V=\pmatrix{0&0&e^{i\delta}&0\cr
 e^{i\alpha}&0&0&0\cr
 0&0&0&e^{i\beta}\cr
 0&e^{i\gamma}&0&0}\,, \end{equation}
\begin{equation} W=\pmatrix{0&e^{i\rho}&0&0\cr
 0&0&0&e^{i\omega}\cr
 e^{i\xi}&0&0&0\cr
 0&0&e^{i\eta}&0}\,. \end{equation}

The reasons for this choice of an 8-parameter unitary matrix
$U$ will
become apparent in the course of the calculation. Some of the
features
can be explained at this stage as follows. We note that,
omitting the
direct-product symbols and replacing unit matrices by 1, etc.,
the
matrix $U$ in (6) has the structure
\begin{equation} 2U=\left(1+\sigma_{zA}\right)V+
 \left(1-\sigma_{zA}\right)W=
V+W+\sigma_{zA}(V-W)\,,\end{equation}
where $V$ and $W$ are operators in the space of $B$ and
$C$. Since
$U$ was chosen diagonal in the space of $A$, the Hamiltonian
$H$ will have a similar structure,
\begin{equation} 2H=P+Q+\sigma_{zA}(P-Q)\,, \end{equation}
with the appropriate $(B\otimes C)$-space Hamiltonians $P$
and
$Q$. Now
in order to avoid three-spin interactions, $P-Q$ must be linear
in
Pauli matrices. On the other hand, we also prefer to avoid
single-spin
(external-field) interactions. Thus, $P+Q$ must contain only
terms of
the second order in the spin components while $P-Q$ must
contain only
terms of the first order in the spin components.
This suggests avoiding putting phase factors on the diagonal,
which
would lead to matrices similar to those encountered in
the NOT-gate calculations$^8$ that are known to be of a
structure undesirable here: they contain a mixture of
first-order and
second-order terms. The off-diagonal choices remaining are
considerably limited; the forms
(7) and (8) are thus nearly unique.

In summary, while the arguments are admittedly vague and they
do
involve a certain level of guess, trial and error, the
presented
parametrization offers a good chance that with further
restrictions
on
the parameters a two-spin interaction Hamiltonian can be
obtained. As
will be seen later, five conditions are imposed so that the
resulting
Hamiltonian depends on three (real) parameters.

\section{The Hamiltonian Matrix}
\noindent
Let us define
\begin{equation} \mu={\alpha+\beta+\gamma+\delta \over 4}\,,  
\end{equation}
\begin{equation} \nu={\rho+\omega+\xi+\eta \over 4}\,,  
\end{equation} 
and also introduce the reduced operators $p$ and $q$
according to
\begin{equation} P=-{\hbar \over \Delta t} p \qquad {\rm and}
\qquad Q=-{\hbar \over \Delta t} q\,. 
\end{equation} 
Then (4) reduces to

\begin{equation} V=\exp (ip) \qquad {\rm and} \qquad W=\exp(iq)\,.
\end{equation}

The following calculations are rather cumbersome. Only the
results will
be presented. The algebraic steps omitted are straightforward.
First,
we diagonalize $V$ and $W$: we calculate their eigenvalues and
also
the matrices of their normalized eigenvectors. The latter can be
used to transform to the diagonal representations.

Specifically, the eigenvalues of
$V$ are $e^{i\mu}$, $ie^{i\mu}$, $-e^{i\mu}$, $-ie^{i\mu}$. The
appropriate eigenvalues of $p$ then follow form (14) as $\mu$,
$\mu+{1\over 2}\pi$,
$\mu+\pi$, $\mu+{3\over 2}\pi$. Arbitrary multiples of $2\pi$ can
be added to these choices. However, there are certain nonrigorous
arguments$^8$ for generally keeping the spread
of eigenvalues of the Hamiltonian as small as possible. Thus,
we choose the simplest expressions.
The eigenvalues of $q$ are determined identically, with $\mu$
replaced
by $\nu$ throughout.

The next step is to apply the inverse of the diagonalizing
transformations for $V$ and $W$ to the diagonal $4\times 4$
matrices
for, respectively, $p$ and $q$. The latter contain the
eigenvalues
 of
$p$ and $q$ as the diagonal elements. The results are the matrix
forms
of the operators $p$ and $q$ in the original representation:

\begin{equation} {4 \over \pi}p=\pmatrix{ {4 \over 
\pi}\mu+3 &-(1+i)
 e^{i(\mu-\alpha)}&-(1-i) e^{i(\delta-\mu)}&-
e^{i(2\mu-\alpha-\gamma)}\cr
-(1-i)e^{i(\alpha-\mu)}& {4 \over \pi}\mu+3 &-
e^{i(2\mu-\beta-\gamma)}&-(1+i) e^{i(\mu-\gamma)}\cr
-(1+i) e^{i(\mu-\delta)}&-e^{i(\beta+\gamma-2\mu)}
& {4 \over \pi}\mu+3 &-(1-i) e^{i(\beta-\mu)}\cr
-e^{i(\alpha+\gamma-2\mu)}&-(1-i) e^{i(\gamma-\mu)}
&-(1+i) e^{i(\mu-\beta)}& {4 \over \pi}\mu+3 }\,, 
\end{equation}

\begin{equation} {4 \over \pi}q=\pmatrix{ {4 \over \pi}\nu+3
&-(1-i) e^{i(\rho-\nu)}&-(1+i) e^{i(\nu-\xi)}
&- e^{i(\rho+\omega-2\nu)}\cr
-(1+i) e^{i(\nu-\rho)}& {4 \over \pi}\nu+3
&- e^{i(\omega+\eta-2\nu)}&-(1-i) e^{i(\omega-\nu)}\cr
-(1-i) e^{i(\xi-\nu)}&- e^{i(2\nu-\omega-\eta)}
& {4 \over \pi}\nu+3 &-(1+i) e^{i(\nu-\eta)}\cr
- e^{i(2\nu-\rho-\omega)}&-(1+i) e^{i(\nu-\omega)}&
-(1-i) e^{i(\eta-\nu)}& {4 \over \pi}\nu+3 }\,. 
\end{equation} 

\section{The Two-Spin-Interaction XOR Hamiltonian}
\noindent
Thus far we decreased the number of independent
parameters in the general unitary transformation and chose it to
be diagonal in the $A$-space. We now ``refine'' our design
of the Hamiltonian to favor interactions of the second
order in the Pauli matrices. First, we note that
both $P$ and $Q$ are constant-diagonal matrices. In terms of
the Pauli matrices, then, both their sum and difference
in (10) will involve constant terms. These are undesirable
because in $\sigma_{zA}(P-Q)$ they lead to terms of order one
(instead
of the desired two), in $H$, while in $P+Q$ they lead to an
additive
constant in $H$ which only affects the overall phase of the
unitary transformation
and is of no interest otherwise. Therefore, we put
\begin{equation} \mu=\nu=-{3\over 4}\pi\,, 
\end{equation} 
in order to nullify these diagonal elements in both $P$ and
$Q$.

Let us now focus our attention on $P-Q$ which, by (17), is now
a matrix with zero diagonal. We must impose conditions on the
parameters to make
$P-Q$ of order exactly one in the Pauli matrices. We note,
however,
that due to zero-diagonal, it cannot contain $\sigma_z$ terms.
The
general form linear in $\sigma_x, \sigma_y$ is

\begin{equation} P-Q={\cal I}_B \otimes \pmatrix{0&X\cr X^*&0}_C+
\pmatrix{0&Y\cr Y^*&0}_B\otimes{\cal I}_C   =\pmatrix{0&X&Y&0\cr
X^*&0&0&Y\cr Y^*&0&0&X\cr 0&Y^*&X^*&0}\,, 
\end{equation}

\noindent where the stars denote complex conjugation, $X$ and $Y$
are
arbitrary (complex) numbers, and $\cal I$ stands for the unit
matrix
as before. Thus we require that $P-Q$ be of the form suggested
by (18). This imposes several rather cumbersome algebraic
conditions: two above-diagonal elements
of the difference must be equal to zero while the remaining four
elements must be equal pairwise. After a lengthy but
straightforward
algebra not reproduced here, we conclude that these conditions
can be satisfied if $\alpha, \beta, \gamma$ are kept as three
independent (real) parameters while the remaining angles are
given by

\begin{equation} \delta=-3\pi-\alpha-\beta-\gamma\,, 
\end{equation}
\begin{equation} \rho=-\pi+\beta\,, 
\end{equation}
\begin{equation} \omega=-2\pi-\alpha-\beta-\gamma\,, 
\end{equation}
\begin{equation} \xi=-\pi+\gamma\,, 
\end{equation}
\begin{equation} \eta=\pi+\alpha\,. 
\end{equation}

These conditions take care of the form of $P-Q$. Interestingly,
our
results below also show that $P+Q$ contains only two-spin
interactions
with this choice of parameters. We have no simple explanation of
this
property (of the absence of first-order terms in $P+Q$). It is
probably related to the fact that the structure pattern of the
original matrices $V$ and $W$ is quite similar even though the
precise positioning of nonzero elements in them is different.
Note that (17) is built into (19)-(23). The explicit
expressions are obtained by a lengthy calculation,

\noindent $P+Q= - {\sqrt{2}\pi \hbar i \over 4 \Delta t} \times$
\begin{equation}
 \pmatrix{0
&e^{-i\alpha}+e^{i\beta}
&e^{-i(\alpha+\beta+\gamma)}-e^{-i\gamma}
&-\sqrt{2}e^{-i(\alpha +\gamma)}\cr
-e^{i\alpha}-e^{-i\beta}
&0
&-\sqrt{2}e^{-i(\beta+\gamma)}
&e^{-i\gamma}-e^{-i(\alpha+\beta+\gamma)}\cr
e^{i\gamma}-e^{i(\alpha+\beta+\gamma)}
&\sqrt{2}e^{i(\beta+\gamma)}
&0
&-e^{-i\alpha}- e^{i\beta}\cr
\sqrt{2}e^{i(\alpha +\gamma)}
&e^{i(\alpha+\beta+\gamma)}-e^{i\gamma}
&e^{i\alpha}+e^{-i\beta}
&0}\,, \end{equation}

\noindent $P-Q= - {\sqrt{2} \pi \hbar i \over 4 \Delta t} \times$
\begin{equation}
\pmatrix{0
&e^{-i\alpha}-e^{i\beta}
&e^{-i(\alpha+\beta+\gamma)}+e^{-i\gamma}
&0\cr
-e^{i\alpha}+ e^{-i\beta}&0&0&
e^{-i(\alpha+\beta+\gamma)}+e^{-i\gamma}\cr
-e^{i(\alpha+\beta+\gamma)}-e^{i\gamma}&0&0&e^{-i\alpha}-
e^{i\beta}\cr 0&-e^{i(\alpha+\beta+\gamma)}-e^{i\gamma}
&-e^{i\alpha}+ e^{-i\beta}&0}\,. \end{equation}

\noindent Finally, we expand these matrices in terms of products of the
Pauli matrices and collect terms according to (10) to get\eject
\noindent $H\,\,=\,\,-{\pi \hbar \over 8 \Delta t} \Big\{ \sqrt{2}
\left( \sin\alpha + \sin\beta \right) \sigma_{zA}\sigma_{xC} 
-\sqrt{2}
\left( \cos\alpha - \cos\beta \right) \sigma_{zA}\sigma_{yC}$\hfill\break

$+ \sqrt{2}\big[ \sin\gamma + \sin(\alpha+\beta+\gamma)
\big] \sigma_{zA}\sigma_{xB} 
- \sqrt{2}\big[ \cos\gamma +
\cos(\alpha+\beta+\gamma) \big] \sigma_{zA}\sigma_{yB}$\hfill\break

$+\sqrt{2}
\left( \sin\alpha - \sin\beta \right) \sigma_{zB}\sigma_{xC} 
-\sqrt{2}
\left( \cos\alpha + \cos\beta \right) \sigma_{zB}\sigma_{yC} $\hfill\break

$- \sqrt{2}\big[ \sin\gamma-\sin(\alpha+\beta+\gamma) \big]
\sigma_{xB}\sigma_{zC} 
+ \sqrt{2}\big[ \cos\gamma -
\cos(\alpha+\beta+\gamma) \big] \sigma_{yB}\sigma_{zC}$\hfill\break

$ -\big[ \sin(\alpha+\gamma)+\sin(\beta+\gamma) \big]
\sigma_{xB}\sigma_{xC}
 +\big[ \cos(\alpha+\gamma)-\cos(\beta+\gamma) \big]
\sigma_{xB}\sigma_{yC} $\hfill\break

$+\big[ \cos(\alpha+\gamma)+\cos(\beta+\gamma) \big]
\sigma_{yB}\sigma_{xC}
+\big[ \sin(\alpha+\gamma)-\sin(\beta+\gamma) \big]
\sigma_{yB}\sigma_{yC}
\Big\}\,. $\hfill $(26)$\break

\noindent Note that (2) corresponds to the parameter choice
$\alpha=\beta=\gamma=0$. The Hamiltonian (26) describes
the three-spin XOR for arbitrary parameter values. All the
interactions involved are two-spin as desired.

The result, however, is not symmetric in any obvious way. It
seems to
correspond to complicated tensor interactions involving
expressions
of order two in the components of the three spins involved. No
rotational or other symmetry in the three-component spin space,
or
planar symmetry, or uniaxial coupling, are apparent. These would
correspond to the familiar Heisenberg, XY, and Ising couplings in
condensed matter physics.
Thus, in order to realize interaction (26) in materials, a
rather
anisotropic medium with highly nonsymmetric tensorial magnetic
interactions will be required.

In this respect our analytical attempt
to ``design'' a multispin quantum gate in this work may indicate
that
different roots to the derivation of Hamiltonians should be also
explored. One could start with the more conventional magnetic
interactions, isotropic (Heisenberg), planar (XY), uniaxial
(Ising),
write down general-parameter Hamiltonians, and then adjust the
coupling parameters numerically in search of those values for
which
useful Boolean gate operations are carried out. There is no
guarantee
that such a program will succeed. We intend to pursue both
approaches in our future work.

In summary, we derived a three-parameter family of Hamiltonians
that correspond to the three-spin XOR gate. While our calculation
demonstrates
that multispin gates can accomplish quantum-logic operations with
two-particle interactions, our results seem also to call
for further work seeking improvement in two ways. Firstly, our
derivation is not general and it has involved a good deal of
guess
work.
Secondly, the terms in the resulting Hamiltonians have no obvious
grouping by symmetries.

\vspace{12pt}\noindent{\tenbf 
Acknowledgements}\vspace{6pt}

\noindent
The work at Clarkson University has been supported in part by US Air
Force grants, contract numbers F30602-96-1-0276 and F30602-97-2-0089. 
The work at Rome Laboratory
has been supported by the AFOSR Exploratory Research Program and by
the Photonics in-house Research Program. This financial assistance 
is gratefully acknowledged.

\vfill\end{document}